\documentstyle[12pt]{article}
\def\Re{{\cal R \mskip-4mu \lower.1ex \hbox{\it e}\,}}
\def\Im{{\cal I \mskip-5mu \lower.1ex \hbox{\it m}\,}}
\def\ie{{\it i.e.}}
\def\eg{{\it e.g.}}

\def\etal{{\it et al.}}

\def\sub#1{_{\lower.25ex\hbox{$\scriptstyle#1$}}}
\def\sul#1{_{\kern-.1em#1}}
\def\sll#1{_{\kern-.2em#1}}
\def\sbl#1{_{\kern-.1em\lower.25ex\hbox{$\scriptstyle#1$}}}
\def\ssb#1{_{\lower.25ex\hbox{$\scriptscriptstyle#1$}}}
\def\sbb#1{_{\lower.4ex\hbox{$\scriptstyle#1$}}}

\def\mh{\ifmmode m\sbl H \else $m\sbl H$\fi}
\def\mch{\ifmmode m_{H^\pm} \else $m_{H^\pm}$\fi}
\def\mt{\ifmmode m_t\else $m_t$\fi}
\def\mc{\ifmmode m_c\else $m_c$\fi}
\def\mz{\ifmmode M_Z\else $M_Z$\fi}
\def\mw{\ifmmode M_W\else $M_W$\fi}
\def\mws{\ifmmode M_W^2 \else $M_W^2$\fi}
\def\mhs{\ifmmode m_H^2 \else $m_H^2$\fi}
\def\mzs{\ifmmode M_Z^2 \else $M_Z^2$\fi}
\def\mts{\ifmmode m_t^2 \else $m_t^2$\fi}
\def\mcs{\ifmmode m_c^2 \else $m_c^2$\fi}
\def\mchs{\ifmmode m_{H^\pm}^2 \else $m_{H^\pm}^2$\fi}
\def\ztwo{\ifmmode Z_2\else $Z_2$\fi}
\def\zone{\ifmmode Z_1\else $Z_1$\fi}
\def\mtwo{\ifmmode M_2\else $M_2$\fi}
\def\mone{\ifmmode M_1\else $M_1$\fi}
\def\tb{\ifmmode \tan\beta \else $\tan\beta$\fi}
\def\xw{\ifmmode x\sub w\else $x\sub w$\fi}
\def\ch{\ifmmode H^\pm \else $H^\pm$\fi}
\def\lum{\ifmmode {\cal L}\else ${\cal L}$\fi}
\def\inpb{\ifmmode {\rm pb}^{-1}\else ${\rm pb}^{-1}$\fi}
\def\infb{\ifmmode {\rm fb}^{-1}\else ${\rm fb}^{-1}$\fi}
\def\epem{\ifmmode e^+e^-\else $e^+e^-$\fi}
\def\ppb{\ifmmode \bar pp\else $\bar pp$\fi}

\def\bsg{\ifmmode b\rightarrow s\gamma \else $b\rightarrow s\gamma$\fi}

\newskip\zatskip \zatskip=0pt plus0pt minus0pt
\def\matth{\mathsurround=0pt}
\def\lsim{\mathrel{\mathpalette\atversim<}}

\def\atversim#1#2{\lower0.7ex\vbox{\baselineskip\zatskip\lineskip\zatskip
  \lineskiplimit 0pt\ialign{$\matth#1\hfil##\hfil$\crcr#2\crcr\sim\crcr}}}

\renewcommand{\thefootnote}{\fnsymbol{footnote}}

\begin{document} \begin{titlepage}
\setcounter{page}{1}
\thispagestyle{empty}
\rightline{\vbox{\halign{&#\hfil\cr
&SLAC-PUB-6455\cr
&February 1994\cr
&T/E\cr}}}
\vspace{0.8in}
\begin{center}

{\Large\bf
$A_{LR}$, Negative $S$, and Extended Gauge Models}
\footnote{Work supported by the Department of
Energy, contract DE-AC03-76SF00515.}
\medskip

\normalsize THOMAS G. RIZZO
\\ \smallskip
{\it {Stanford Linear Accelerator Center\\Stanford University,
Stanford, CA 94309}}\\

\end{center}

\begin{abstract}

The implications of the recent measurement of the left-right asymmetry,
$A_{LR}$, by the SLD Collaboration for theories with extended gauge
sectors is examined.  We show that it is possible to arrange for large,
negative values of $S$, based on an analysis of leptonic data, without
serious side effects for other observables in certain classes of models. The
implications of such scenarios for future measurements on the $Z$ peak, at the
Tevatron, and for atomic parity violation  experiments are examined.

\end{abstract}

\vskip0.30in
\begin{center}

Submitted to Physical Review {\bf D}.

\end{center}


\renewcommand{\thefootnote}{\arabic{footnote}} \end{titlepage}


The left-right polarization asymmetry, $A_{LR}$, provides one of the most
sensitive probes of the standard model (SM) leptonic couplings of
the $Z$ boson. At tree level, $A_{LR}$ is directly related to the ratio of
the vector to the axial vector coupling of the electron and is independent of
the nature of the fermions produced in the final state, \ie,
\begin{equation}
A_{LR}={2y\over {1+y^2}} \,,
\end{equation}
where $y=v_e/a_e=1-4x_{eff}$ in the SM. Recently, the SLD
Collaboration{\cite {sld}}  has announced a new, high-precision determination
of $A_{LR}$ based on their 1993 data sample,
$A_{LR}=0.1656\pm 0.0073\pm 0.0032$, which
should be compared to the earlier result{\cite {oldsld}} from their 1992 data,
$A_{LR}=0.100\pm 0.044\pm 0.004$. (Both determinations essentially make
use of only the
hadronic final states in $e^+e^-$ annihilation.) Within the SM, box and
vertex corrections can be directly evaluated so that the radiatively corrected
values for $x_{eff}$ can be extracted from both determinations:
$x_{eff}= 0.2288\pm 0.0009\pm 0.004 ~(1993)$
and $x_{eff}=0.2378\pm 0.0055\pm 0.0005 ~(1992)$ which when folded
together yield
$x_{eff}=0.2290\pm 0.0010$ if the errors are combined in quadrature. The
surprise here is that this value of $x_{eff}$ is quite different than that
obtained from LEP data alone, $x_{eff}=0.2324\pm 0.0005$, or when LEP data
is combined with $W$ boson mass determinations and low energy neutral current
data, $x_{eff}=0.2325\pm 0.0005${\cite {lep}}, both of which are several
$\sigma$ away from the SLD value. The implications of this apparent conflict
are at the moment unclear as the LEP value is `supported' by the latest set
of preliminary $W$ mass measurements by both CDF and D0{\cite {kek}}
(which now yield a new
world average value of $M_W=80.21\pm 0.18$ GeV).
On the
otherhand, the $A_{LR}$ measurement is extremely sensitive to $x_{eff}$, is
exceptionally clean, and has totally different systematics than the LEP
experiments. (We note in passing that for top-quark masses in the range
160-180 GeV, near the central value extracted from the radiative corrections
analyses of LEP data,
$x_{eff}$ is numerically equal to $sin^2 \theta_{\overline {MS}}(M_Z)$ to
a very high precision, as defined in the
minimal way{\cite {sir}}, \ie, when the additional Marciano-Rosner subtraction
scheme{\cite {mr}} is not employed.)

One approach is to combine the $A_{LR}$ measurement with the other existing
data and see what we can learn. If we take the combined LEP
asymmetry and leptonic
width results together with the latest $M_W$ determination and the value
of $A_{LR}$ (as
they are all rather insensitive to the value of $\alpha_s(M_Z)$), a
Peskin-Takeuchi{\cite {pt}} type analysis can be performed which yields the
central values $T\simeq -0.38\pm 0.34$ and $S\simeq -0.58\pm 0.30$ assuming
$m_t=165$ GeV
and $M_{Higgs}=300$ GeV{\cite {peskin}}. In what follows, we will adopt these
values as input into our analysis as they conform to the most recent complete
fit to electroweak data performed in Ref.{\cite {lep}}.
(For larger values of the top-quark
mass, $m_t=175$ GeV, say,  the central value for $T$ decreases by about 0.20
and $S$ increases only very slightly for fixed $M_{Higgs}$. $S$ and $T$ would
then have comparably negative values which are about $2\sigma$ from zero.)
We should note, however, that in the {\it absence}
of the new $A_{LR}$ result from SLD, both $S$ and $T$ would be essentially
zero as we might
have anticipated based on the well-known excellent fit of the SM to
the previously existing data. The SM is certainly very far from being
excluded by this new analysis (as the SM prediction still lies well within the
$90\%$ CL ellipsoid), however, we are led to contemplate what kind of new
physics would push the fit for the leptonic data
down into the negative $S,T$ quadrant without much
influence elsewhere.

In this paper, we examine several extended electroweak gauge models, which
predict the existence of a new $Z'$ (and in some cases, a $W'$) gauge
boson(s), in the light of the recent $A_{LR}$ measurement by SLD.
In particular,
we are interested in identifying models that move us closer to the central
values of $S$ and $T$ above without overly disturbing other electroweak
observables such as $R_{inv}=\Gamma_{inv}/\Gamma_{\ell}$,
$R_h=\Gamma_{had}/\Gamma_{\ell}$, $R_b=\Gamma_b/\Gamma_{had}$, $M_W$, and
$A_b$, the b-quark asymmetry parameter. The present analysis
only uses these separate
quantities as additional constraints on the parameters of
potential $Z'$ models. We remind the
reader that the constraints on the existence of a new $Z'$ from radiative
corrections analyses made before the announcement of the SLD result were
very strong{\cite {rad}}, in some cases requiring $M_{Z'}>0.5-1$ TeV.
This merely
reflected the observation that since the SM fit the data so well there was
little room remaining for significant shifts in observables due to
new physics. The new SLD result for $A_{LR}$, now provides a bit more
breathing space for a relatively light $Z'$, possibly in the mass range
that can be explored at the Tevatron in the future, \eg, 500-800 GeV.
In this analysis, we are
trying to identify if regions exist in a given extended gauge model's
parameter space that allow large, negative $S$ (as defined by the
leptonic data) without upsetting the values of the other observables. These
regions, if they exist for a given model, will be ones where $v_e$ experiences
a significant increase in magnitude while the other $Z$ couplings
will be little affected. The `successful' regions of parameter space we
find below are meant only to be {\it suggestive} as they depend upon the
specific values of the input parameters, \eg , $m_t$, that we employ in
this analysis. As we will see below, it is not always possible for models to
produce a significant shift in $v_e$ without there being sizeable changes in
the other couplings.

Direct searches for a $Z'$ at the Tevatron by CDF have resulted in a
preliminary lower limit of
495 GeV, from a partial analysis of the electron data from the 1992-93
run Ia{\cite {cdfzp}}, {\it assuming} SM couplings and that the $Z'$ decays
to SM particles only. We note that the full analysis of the data from
this run, including muons, may increase this limit by about 90 GeV. In
addition, run Ib has already commenced which will increase these limits
even further. In any of the more realistic extended electroweak models(EEMs),
the $Z'$ couplings are sufficiently different
from those of a SM $Z$ so that the actual mass limits could be significantly
higher or lower than the quoted 495 GeV result. For example, in
$E_6$ models(ER5M){\cite {us}}, under identical assumptions, the
corresponding bounds hover near 400-420 GeV for all values of the $E_6$ mixing
parameter $-\pi/2 \leq \theta \leq \pi/2$ while in the Left-Right
Symmetric Model(LRM){\cite {lrm}}
the limit is 445 GeV assuming the ratio of $SU(2)_R$ to $SU(2)_L$ gauge
couplings, $\kappa=g_R/g_L=1$. Similarly, in the Alternative LRM of Ma \etal
{}~(ALRM){\cite {alrm}}, one finds a $Z'$ mass limit of nearly 550 GeV
but in the Un-Unified Model(UUM){\cite {harv}} of Georgi \etal, the limits
vary from 400 to 600 GeV depending upon the value of the model parameter,
$0.22 \leq s_{\phi} \leq 0.99$. As we will see below, it is
quite easy for the $Z'$'s of interest
to us to satisfy these direct search constraints. We will assume for
simplicity
that in models where a $W'$ is present, it plays no role in low-energy
processes and does not mix with the SM $W$. We note that while the above
list of extended models is reasonably representative it is far
from exhaustive as the
literature on this class of extensions to the SM is quite robust. For the
explicit expressions of the various fermion couplings in each of the models
described above, we refer the reader to the original literature.

We first wish to explore these various models to find out which,
if any, have parameter spaces that allow us to move toward the negative $S$
and $T$ regions discussed above. While a general Peskin-Takeuchi analysis
{\it cannot} be applied to an extended gauge model as a whole, since the
$Z'$ can induce significant
flavour-dependent modifications in fermionic couplings, a restricted
analysis of this kind is possible if we limit ourselves to leptonic
observables at the $Z$-pole and $M_W$. The reasoning here is clear; since
there are only three observables under consideration one is completely free
to parameterize their potential deviations from SM predictions in terms of
three variables which can be identified as $S,T$ and $U$.
This has been pointed out most clearly by the work of Altarelli
\etal {\cite {rad2}}, and this particular approach has been employed by
other authors to constrain some other extended models{\cite {tau}} not
discussed in the present work. This
particular set of observables has the added
advantage of being quite insensitive to the precise value of $\alpha_s(M_Z)$.
The influence of a $Z'$ and small $Z-Z'$ mixing (through an angle $\phi$) has
three direct effects on these observables which can be summarized
as follows. Due to mixing, the $Z$ and $Z'$
form the mass eigenstates $Z_{1,2}$ with $M_1<M_Z$, the SM $Z$ mass. However,
using the {\it observed} mass (\ie, $M_1$) as an input parameter ($i$)
modifies the traditional $W-Z$ mass relation by introducing an effective
$\rho$-parameter, $\rho=1+\delta \rho$, where $\delta \rho$ is of order
$r=M_1^2/M_2^2$ in models where $SU(2)_L$ breaking is performed solely by
isodoublets(as will be the case for all the models we examine below). This
$\rho$ parameter also produces an overall rescaling of the
$Z$ partial widths, as calculated in the `$G_F$'-scheme, when the measured
$Z_1$ mass is used as input. ($ii$) The SM vector
and axial-vector couplings $v,a$ are directly modified by the small
admixture of the corresponding $Z'$ couplings $v',a'$. ($iii$) If one uses
the observed $Z_1$ mass to define the weak mixing angle, the shift from the SM
$Z$ mass due to mixing induces a corresponding
change in the value of $sin^2 \theta_w$ that should be employed in the
evaluation of fermionic couplings.

In almost all models of interest, including those discussed here, $\phi$,
$r$, and $\delta \rho$ are directly related to each other via a
model-dependent parameter,
$\gamma$, which is of order unity and is sensitive to the details of the
symmetry breaking scheme of the extended model. In terms of the elements
of the $Z-Z'$
mass matrix, $\gamma$ is defined by writing the matrix in the form
\begin{eqnarray}
{\cal M}^2 & = & \left( \begin{array}{ccc}
M_Z^2 & \gamma M_Z^2 \\
\gamma M_Z^2 & M_{Z'}^2
\end{array} \right) \,,
\end{eqnarray}
which exploits the fact that the vacuum expectation values
(vev's) contributing to both the ${\cal M}_{11}^2$ and ${\cal M}_{12}^2$
elements are the
same. The particular values of $\gamma$ for the models
above have been
discussed in detail elsewhere{\cite {old}} and we simply give the relevant
expressions below:
\begin{eqnarray}
\gamma_{LRM} & = & -(\kappa^2-(1+\kappa^2)x_w)^{1/2} \,, \nonumber \\
\gamma_{ALRM} & = & {x_wt^2_\beta-(1-2x_w)\over (1-2x_w)^{1/2}(1+t^2_\beta)}
\,, \\
\gamma_{ER5M} & = & -2\sqrt{{5x_w\over 3}}\Bigl[\Bigl( {c_\theta\over\sqrt 6}
-{s_\theta\over\sqrt 10}\Bigr) t^2_\beta - \Bigl( {c_\theta\over\sqrt 6}
+{s_\theta\over\sqrt 10}\Bigr)\Bigr] (1+t^2_\beta)^{-1} \,, \nonumber \\
\gamma_{UUM} & = & {-(1-x_w)^{1/2}s_{\phi}\over{(1-s_{\phi}^2)^{1/2}}} \,,
\nonumber
\end{eqnarray}
where $x_w=sin^2 \theta_w$, $t_\beta=\tan\beta=v_t/v_b$, the usual ratio
of vacuum expectation
values responsible for the top and bottom quark masses, and $s_\theta(
c_\theta)=\sin\theta(\cos\theta)$ being the sine and cosine of the
ER5M mixing angle discussed above.
Note that if $\theta=-90^\circ$ (model $\chi$) then $\gamma_\chi=
-(2x_w/3)^{1/2}$ is independent of the value of $\tan\beta$.
To lowest order in $r$, one then finds the simple result
\begin{eqnarray}
\phi & = & -\gamma r \,, \\
\delta \rho & = & \gamma^2 r \,. \nonumber
\end{eqnarray}
To similar leading order in  $\phi$ (or $r$), Altarelli \etal {\cite {rad2}}
{}~then obtain the following relations for the shifted values of $S,T$, and
$U$:
\begin{eqnarray}
\Delta T & \simeq & \alpha^{-1}(\delta \rho -4a'\phi) \,, \nonumber \\
\Delta S & \simeq & 2\phi \alpha^{-1}[(1-2x)v'-(1+2x)a'] \,, \\
\Delta U & \simeq & 4\phi \alpha^{-1}(v'+3a') \,, \nonumber
\end{eqnarray}
where $x$ is simply the value of $sin^2 \theta_w$ one would obtain in the SM
limit, which we take to be 0.2325 in light of the discussion above and the
assumed values of $m_t$ and $M_{Higgs}$ we use as input into this analysis.
$\alpha^{-1}\simeq 128$ and
$v'$ and $a'$ are the charged lepton couplings to the $Z'$ normalized as
in SM.
Of course, in the results presented below, we use only exact expressions which
include all the higher order terms in $\phi$ (or $r$) and not the suggestive
approximate forms given above. These approximate expressions do,
however, reproduce the exact results at the level of $5\%$ or so for the
cases of interest and show us precisely which combinations of the
properties of the $Z'$ are being probed. The exact expressions are cumbersome
and not very enlightening and thus we do not reproduce them here.

To demonstrate that an arbitrary extended model will
not put us into the $S-T$ range of
interest, we first consider the case of the ALRM. The couplings in this model
are free of independent parameters and $\gamma$ depends solely on the ratio of
the two Higgs doublet vev's (which as mentioned above is traditionally
denoted by $tan \beta$) that are present in the model. The only
additional parameter we need to consider is the mass of the $Z'$ itself.
As $tan \beta$ is varied in this model, for fixed $M_{Z'}$, a curve is
traced out in the $S-T$ plane as is
shown in Fig.~1. Here we see that this model populates the wrong part of the
$S-T$ plane and never reaches sufficiently large negative values of $S$,
close to the central value of the leptonic $S-T$ fit described above. This
model demonstrates that it is not obvious that a given extended model
can actually
produce the desired range of $S,T$. In fact, many other extended models tend
to favour $S>0$, an example of which is the universality violating
model discussed in Ref. {\cite {tau}}.

However, some models can produce negative values of $S$ and $T$
(especially if they have greater
parameter flexibility) an excellent example being those based on the gauge
group $E_6$, \ie , the Effective Rank-5 Models(ER5M). In Figs. 2a and
2b, we present the values of $S$ and $T$ as functions of the $E_6$ parameter
$\theta$ assuming a representative $Z'$ mass of 750 GeV and a range of
$tan \beta$ values.  Here we see that once $tan \beta$ exceeds 5-10 the curves
become quite indistinguishable.
Figs. 3a and 3b show the resulting $S-T$ plots for this model assuming
$M_{Z'}=500$ and 750 GeV, respectively, for the same set of $tan \beta$
values. As $\theta$ is varied, the curves form closed ellipses which share
a common point at $\theta=\pm 90^\circ$ where the results are $tan \beta$
independent. Also shown on these figures is the `data point' corresponding
to the $S-T$ fit described above. Similarly, Fig. 3c shows the case with
$tan \beta=20$ held fixed but with $Z'$ masses varying between 500 and 1500
GeV. For low masses, it is clear that sufficiently negative values
of $S$ and $T$ are
easily reached but this becomes more difficult as $M_{Z'}$ increases
beyond $\sim 1$ TeV. In fact, for $M_{Z'}=500(600, 750, 850)$ GeV,
the best fit is provided by $\theta=24^\circ(19.5^\circ, 9.5^\circ,
0.5^\circ)$ with
the corresponding $tan \beta$  values of 3(4, 8.5, 100). ($tan \beta=100$ was
assumed to be the maximum allowed value but the difference in the fit between
$tan \beta=20$ and 100 is very minimal.) For larger masses, the best fit value
for $\theta$ becomes
negative (as we are pushed to the lower left end of the ellipse's major axis
corresponding to increasingly negative values of $\theta$)
while $tan \beta$ assumes its maximally allowed value, hence, the
choice of a large $tan \beta$ in Fig. 3c.  These best fit values are
suggestive of the region of the model's parameter space that is preferred by
the $S-T$ analysis. We thus conclude that the ER5M
with a $Z'$ in the 500-850 GeV can provide reasonable $S,T$ values but larger
masses would have somewhat higher $\chi^2$. This entire mass range is clearly
accessible at the Tevatron for integrated luminosities greater
than 100-500 $pb^{-1}$.

Interestingly, the regions in the ER5M parameter space which yield negative
values for both $S$ and $T$ near the central values of the fit lead to
very small fractional changes in all of the SM $Z$ fermionic
couplings {\it except} for the electron's vector coupling. Specifically, for
the case of a 750 GeV $Z'$ with $\theta$ and $tan \beta$ values in the
neighborhood of the ranges quoted
above, we typically find that $v_d, a_d, v_b, a_b$, and $a_e$ are only
modified at the level of $0.2\%$, $a_u$ and $v_{\nu}=a_{\nu}$ by $0.5\%$,
and $v_u$ by $0.7\%$. Shifts of similar magnitude are also
encountered for the lower mass 500 and 600 GeV $Z'$ cases assuming the
specific values of $\theta$ and $tan \beta$ listed above.

We must be sure to check that for the above parameter choices,
other electroweak observables are not overly affected or perhaps
lead to improvements
in comparison to the data since combinations of the the small
changes in the individual couplings may conspire together to cause a
significant deviation.  Based on the apparent shifts in the couplings
discussed in the previous paragraph, however, we expect to be quite safe.
To prove that this is indeed the case, we show the ratio of
the predictions of the ER5M to those of the SM
for $R_h, R_b, R_{inv}, A_b$ as well as the predicted
fractional shift in $M_W$ in Figs. 4a-e as a function of $\theta$ with
$M_{Z'}=750$ GeV.
In order to be specific, we will assume $\alpha_s(M_Z)=0.123$ when performing
our numerical evaluations{\cite {lep}}. For
values of $\theta$ (or $tan \beta$) sufficiently far away from the range
which yields negative $S$ and $T$ as discussed above, we see that
significant shifts may occur in any or all of
these observables. However, for the specific range of parameters of interest
to us very little influence from $Z-Z'$ mixing is noted. Typically,
the largest deviations we find
are an increase in $R_{inv}$ by $\simeq 0.67\%$ (\ie, $\Delta N_{\nu}=
0.0022$), a decrease in $R_b$ by $\simeq 0.39\%$, an increase in $R_h$ by
$\simeq 0.4-0.5\%$, and an upward shift in
$M_W$ by about 80 MeV, all of which are at the level
being probed by current experiment. The existing $95\%$ CL upper limits on
the allowed  variation in these quantities (in the directions that they are
shifted within the model) are
approximately $1.39\%, 1.34\%, 0.47\%$ and 350 MeV,
respectively{\cite {lep,kek}}, for $m_t=165$ GeV and $M_{Higgs}=300$ GeV
with $\alpha_s(M_Z)=0.123$ held fixed. Shifts of similar magnitude are
also found for lighter (and somewhat heavier) $Z'$ masses for
parameters tuned near the above choices.
We can thus conclude that the ER5M has sufficient freedom such that
for a reasonable range of $Z'$ masses (500-1000 GeV, say) we can find values of
the parameters $\theta$ and $\tan \beta$ that lead to large negative $S$,
through a significant shift in $v_e$, without similar drastic changes in the
other couplings or direct observables. However, the predicted shifts are not
so small as to render them unobservable in the near future and the $Z'$ masses
are not so large as to make direct production of a new neutral gauge boson
arising from this model impossible to observe at the Tevatron.

Next, we turn our attention to the LRM with the variable $\kappa=g_R/g_L$
as the only free parameter in addition to $M_{Z'}$. As such, there is
clearly much less freedom in
the model. Fig. 5 shows the $S-T$ plane for this
model assuming the
$Z'$ mass lies in the 500-1500 GeV range; possible effects from the $W_R'$ are
ignored. The `curves' are essentially
straight lines that penetrate into the negative $S$, negative $T$
quadrant as the value of $\kappa$ is varied. As $\kappa$ is increased, we
move further down and to the left along the curve for each value of the $Z'$
mass. Clearly, somewhat heavier ($>750$ GeV) $Z'$ masses are favored by the
$S-T$ region fit to the present data. For $0.55 < \kappa <2$, we find that
a $Z'$ in the
0.8-3.0 TeV mass range will yield results for $S$ and $T$ quite close to the
central values from the fit. (We restrict our attention to this $\kappa$ region
as we expect on general grounds that this ratio should not be too
different from unity as suggested by grand unified models. Finiteness of the
$Z'$ couplings in this models also requires that
$\kappa^2 > {x_w\over {1-x_w}} \simeq 0.55$.)
As $M_{Z'}$ increases, the best fit values of
$\kappa$ also increases so that for larger masses, the restricted range
of $\kappa$ we employ is insufficient to reach close to the $S-T$ central
values. For a $Z'$ mass of 800(1000, 1200, 1500, 2000, 3000) GeV, the range
of $\kappa$ values with the best $\chi^2$'s is centered correspondingly at
0.82(0.89, 0.98, 1.13, 1.39, 1.95).  As in the ER5M case, we must also test
that these values of the model parameters do not significantly modify the
other observables. (We still will implicitly assume that the existence
of the $W'$ has no influence here.) Figs. 6a-e show the $\kappa$ sensitivity
of the observables discussed above all of which are found to slightly increase
in magnitude in comparison to their SM predictions. When the best fit values
of the $\kappa$'s are employed we see that
all the variations are safely small provided $M_{Z'}$ is greater than
about $\sim 1$ TeV or so. Thus for $Z'$ masses in the 1-3 TeV range with
the above
$\kappa$ values the LRM will yield only small modifications to nonleptonic
observables and will still produce negative values of $S$ and $T$ in the
range of interest.

The last model we consider is the UUM where we will again assume for
simplicity that $W'$ effects can be ignored. The mass matrix parameter
$\gamma$ in this model is completely determined by the value of $s_{\phi}$
so there is less parameter freedom than in the $E_6$ ER5M scenario.
Fig. 7 shows the $S-T$ plane
for this model where the general behaviour $S \simeq T<0$ is observed. (This
would be a particularly nice prediction if the top-quark mass were 175 GeV as
discussed above.) Here,
the curves for different $Z'$ masses lie atop one another and as
the free parameter $s_{\phi}$ increases we move away from the region of the
origin out towards negative $S,T$. We arrive near the central part of the
$S,T$ region of
interest with $s_{\phi}$=0.61(0.76, 0.84, 0.89, 0.92) assuming $M_{Z'}$=
1(1.5, 2, 2.5, 3) TeV.  However, even for sizeable $Z'$ masses we find that
some of the other observables are significantly altered. While the resulting
shifts in $R_b, A_b$, and $R_{inv}$ are found to be quite small,
below $\sim0.1\%$, in
all the cases above, we find that $R_h$ is significantly increased by
$1.1-1.7\%$, with the magnitude of the shift decreasing very slowly with
larger $Z'$ mass.
Similarly, the $W$ mass is shifted upwards by about 160 MeV in all cases as
larger values of $M_{Z'}$ are compensated for by the correspondingly
increasing values of $s_{\phi}$ required by the $S-T$ analysis. While
we might be able to defeat any potential $W$ mass shift problem by
allowing a small $W-W'$ mixing, the
rather large increase in $R_h$ is too big to be tolerated by
existing data even when we allow for the uncertainty in $\alpha_s(M_Z)$.
To reduce the upward shift in $R_h$ to a manageable level,
below $\lsim 0.4\%-0.5\%$, would require increasing the $Z'$ mass to
the multi-TeV
range (beyond what can be easily probed by the LHC) and fine-tuning
$s_{\phi}$ to values extremely close to unity. This would force the $Z'$ to be
strongly coupled as discussed in the last two papers in Ref.{\cite {harv}}.
Thus, unless we allow for an extremely massive, strongly coupled  $Z'$
and (possibly) significant $W-W'$ mixing, the UUM does not
provide adequately for the possible shift in the $S,T$ parameters while
leaving other observables essentially unaffected.

Besides the $W$ mass measurement, direct $Z'$ production, and the refinement
of the data for on-resonance observables, how can we probe the physics of
these models above in the near future? One possibility is to
further improve the measurements{\cite {qw}} of the weak charge, $Q_W$, as
determined by atomic parity violation experiments. The sensitivity of $Q_W$
to the existence of a $Z'$ has been discussed in the recent literature by
several authors{\cite {qwt}}. As above, we take the SM value of $Q_W$ to be
that given by the choice $m_t=165$ GeV and $M_{higgs}=300$ GeV corresponding
to $sin^2 \theta_{\overline {MS}}(M_Z)=0.2325$. For Cesium, the current
experimental value of
$Q_W$ is $-71.04\pm 1.58\pm 0.88$, the SM predicts $-73.25$, yielding
$\Delta Q_W=Q_W^{exp}-Q_W^{SM}=2.21\pm 1.81$. Future
experiments are expected to reduce these errors by a factor of order 5-6 making
such measurements competitive with $M_W$ determinations in probing electroweak
corrections.

How large and what sign is the predicted shift in the value of $Q_W$,
$\delta Q_W$, due to the existence of a $Z'$ for the models discussed above?
(We note in passing that the {\it effective} value of the $S$ parameter
extracted from atomic parity violation measurements in Cesium is essentially
given by $S_{eff}=-\delta Q_W/0.795$.)
Figs.~8a-b show the $E_6$ model predictions for a $Z'$ of mass 500 or 750
GeV, respectively, as a function of $\theta$, assuming different $tan \beta$
values. For the 500 GeV case,
we see $\delta Q_W$ is very small and positive near $\theta=24^\circ$
if $tan \beta=3$, while for the corresponding 750 GeV example,
$\delta Q_W \simeq 0.25$. For the
LRM scenario, Fig.~9 shows that $Z'$ masses in the 1-2.5 TeV range yield small,
positive predictions for $\delta Q_W \simeq 0.2$ when the
values of $\kappa$ found above are
used. Thus we see a general pattern arising, which suggests that our
`successful'
models predict small, positive increases in $Q_W$ (\ie , small, negative
values of $S_{eff} \simeq -0.3$) which should be observable
during the next round of atomic parity violation experiments.

In this paper, we have examined the possibility that extended electroweak
gauge models can allow for a large and negative value of $S$, as determined
by LEP/SLC leptonic decay and asymmetry data, while not significantly
affecting other observables. What is required is a region of the model's
parameter space where the charged lepton vector coupling, $v_e$,
is significantly
increased while the corresponding fractional deviations in all other couplings
are obliged to remain small. This requirement is far from trivial and cannot
be realized in most models; even the `successful' models only do so over a
relatively narrow range of parameters.

Specifically, our results can be summarized as follows:

($i$) The Alternative Left-Right Model was found to lead to values of $S$
and $T$ which populated the
wrong regions of the $S-T$ plane, \eg , when $S$ was sufficiently negative,
$T$ was very large and positive.

($ii$) The Un-unified model, while easily obtaining $S,T$ values of interest
over a wide range of $Z'$ masses,
always resulted in too large an increase in $R_h$ by a factor of 2-3 beyond
what the existing data can tolerate. This was found to be true even
for extremely large
$Z'$ masses. This situation {\it might} be avoided if the model parameter
$s_{\phi}$ were tuned extremely close to unity but then we would pay the
price of having extreme fine-tuning and a strongly coupled new gauge sector.

($iii$) The Left-Right Model was found to easily satisfy all of the necessary
constraints for $Z'$ masses in the 1-3 TeV range with appropriately chosen
values of the parameter $\kappa=g_R/g_L$, {\it assuming} that the $W'$ did
not influence low energy physics. These $Z'$'s would clearly be beyond the
range accessible to the Tevatron and must await discovery at the LHC and NLC.

($iv$) The $E_6$ models, which perhaps have the most flexibility amongst those
models examined here, were also found to be able to satisfy all the necessary
conditions for $Z'$ masses in the 500-1000 GeV range for values of the model
parameters $0^\circ \leq \theta \leq 24^\circ$ and $tan \beta >3$. This
$Z'$ mass
range {\it is} accessible at the Tevatron for integrated luminosities in
excess of 100 $pb^{-1}$ which may be achieved in the not too distant
future.

($v$) In addition to $Z$-pole, $W$-mass, and direct $Z'$ searches, perhaps
one of the best signatures for the models discussed here is a small positive
increase in
the value for the weak charge in Cesium, $\delta Q_W \simeq 0.2-0.3 $, in
comparison to the SM prediction. In the usual language, this would correspond
to extracting an effective value of the $S$ parameter from these measurements
in the range $-0.25$ to $-0.4$. Such
a shift was observed to occur for both the LRM and $E_6$
models in the parameter ranges of interest. Future experiments searching for
atomic parity violation should be sensitive to such effects.

Hopefully precision measurements may soon begin to yield some evidence for
new physics beyond the Standard Model.

\vskip.25in
\centerline{ACKNOWLEDGEMENTS}

The author would like to thank J.L.\ Hewett, M.\ Peskin, M.\ Swartz, and
E.\ Nardi for discussions related to this work. The author would also like
to thank the
members of the Argonne National Laboratory High Energy Theory Group for use of
their computing facilities.

\newpage

%
\def\MPL #1 #2 #3 {Mod.~Phys.~Lett.~{\bf#1},\ #2 (#3)}
\def\NPB #1 #2 #3 {Nucl.~Phys.~{\bf#1},\ #2 (#3)}
\def\PLB #1 #2 #3 {Phys.~Lett.~{\bf#1},\ #2 (#3)}
\def\PR #1 #2 #3 {Phys.~Rep.~{\bf#1},\ #2 (#3)}
\def\PRD #1 #2 #3 {Phys.~Rev.~{\bf#1},\ #2 (#3)}
\def\PRL #1 #2 #3 {Phys.~Rev.~Lett.~{\bf#1},\ #2 (#3)}
\def\RMP #1 #2 #3 {Rev.~Mod.~Phys.~{\bf#1},\ #2 (#3)}
\def\ZP #1 #2 #3 {Z.~Phys.~{\bf#1},\ #2 (#3)}
\def\IJMP #1 #2 #3 {Int.~J.~Mod.~Phys.~{\bf#1},\ #2 (#3)}

\newpage

%
{\bf Figure Captions}
\begin{itemize}

\item[Figure 1.]{$S-T$ plot for the ALRM assuming $Z'$ masses of 500(dots),
750(dashes), 1000(dash-dots), 1250(solid), or 1500(square-dots) GeV. The
value of the parameter $tan \beta$ varies along each curve. The $S,T$ origin
was assumed to correspond to $m_t=165$ GeV and $M_{Higgs}=300$ GeV within
the SM.}
\item[Figure 2.]{Values of the parameters (a)$T$ and (b)$S$ in the ER5M case as
functions of the parameter $\theta$ assuming $M_{Z'}=750$ GeV. From top to
bottom, the curves correspond to $tan \beta=1, ~1.5, ~2, ~3, ~5, ~10, ~40$.}
\item[Figure 3.]{$S-T$ plots for the ER5M assuming a $Z'$ mass of (a)500 or
(b)750 GeV for the same set of $tan \beta$ values as in Fig.~2. The plotted
`data' point corresponds to the $S-T$ fit to the LEP and SLC leptonic data.
{}From right to left, the ellipses correspond to $tan \beta=1, ~1.5, ~2, ~3,
{}~5,
{}~10, ~40$. (c) Same as (a) and (b) but with $tan \beta=20$  for increasing
$Z'$ masses of 500(dots), 750(dashes), 1000(dash-dots), 1250(solid), and
1500(heavy solid) GeV.}
\item[Figure 4.]{The ratio of the predicted values for (a)$R_h$, (b)$R_b$,
(c)$R_{inv}$, and (d)$A_b$ in the ER5M compared to the SM for a 750 GeV $Z'$.
(e)The corresponding fractional shift in the $W$ mass. The curves are for
the same values of $tan \beta$ as shown in Figs.~2a-b, with the smallest
value of $tan \beta$ corresponding to the lowest dotted curve.}
\item[Figure 5.]{$S-T$ plot for the LRM for the same $Z'$ mass values
displayed  in Fig.~3c together with the `data' point representing the $S,T$
fit.}
\item[Figure 6.]{Same as Fig.~4, but for the LRM as a function of the
parameter $\kappa$. From top to bottom the results shown are for $Z'$ masses
of 0.8, 0.9, 1, 1.2, 1.5, and 2 TeV.}
\item[Figure 7.]{$S-T$ plot for the UUM; the curves corresponding to the
different $Z'$ masses discussed in the text ($1-3$ TeV) lie atop one another.}
\item[Figure 8.]{$\delta Q_W$ as predicted in the $E_6$ model case as a
function of $\theta$ for a $Z'$ mass of (a) 500 and (b) 750 GeV. From bottom
to top, the curves on the right-hand side of the figure correspond to the same
values of $tan \beta$ as in Fig.~2.}
\item[Figure 9.]{Same as Fig.~8, but for the LRM as a function of $\kappa$.
{}From top to bottom, the curves correspond to a $Z'$ mass of 1, 1.25, 1.5,
1.75,
2, 2.25, and 2.5 TeV, respectively.}
\end{itemize}


\begin{thebibliography}{99}
\bibitem{sld}
B.\ Schumm, SLD Collaboration, talk given at the {\it Tenth Aspen Winter
Physics Conference: Particle Physics Before the Year 2000},
Aspen, Colorado, January, 1994.
\bibitem{oldsld}
K.\ Abe \etal , SLD Collaboration, \PRL 70 2515 1993 .
\bibitem{lep}
The LEP Collaborations, CERN report CERN/PPE/93-157, 1993.
\bibitem{kek}
See talks by D.\ Saltzberg, CDF Collaboration, and Q.\ Zhu, D0 Collaboration,
given at the {\it Ninth Topical
Workshop on Proton-Antiproton Collider Physics}, Tsukuba, Japan, October 1993.
\bibitem{sir}
P.\ Gambino and A.\ Sirlin, \PRD D49 1160 1994 .
\bibitem{mr}
W.J.\ Marciano and J.L.\ Rosner, \PRL 65 2963 1990 .
\bibitem{pt}
M.E.\ Peskin and T.\ Takeuchi, \PRL 65 964 1990 ~and \PRD D46 381 1992 .
\bibitem{peskin}
The author would like to thank Michael Peskin for providing him with the
results of this analysis.
\bibitem{rad}
See, for example, W.\ Hollik and M.\ Swartz, talks given
at the {\it XVI International
Symposium on Lepton-Photon Interactions}, Cornell University, August 1993.
\bibitem{cdfzp}
F.\ Abe \etal, CDF Collaboration, Fermilab report Fermilab-Pub-93-211, 1993.
\bibitem{us}
J.L.\ Hewett and T.G.\ Rizzo, \PR 183 193 1989 .
\bibitem{lrm}
For a review of the LRM and original references, see R.N. Mohapatra, {\it
Unification and Supersymmetry}, (Springer, New York, 1986).
\bibitem{alrm}
E. Ma, \PRD D36 274 1987 ~and \MPL A3 319 1988 ;
K.S. Babu \etal , \PRD D36 878 1987 ;
V. Barger and K. Whisnant, \IJMP A3 879 1988 ;
J.F. Gunion \etal , \IJMP A2 118 1987 ;
T.G. Rizzo, \PLB B206 133 1988 .
\bibitem{harv}
H. Georgi, E.E. Jenkins, and E.H. Simmons, \PRL 62 2789 1989 ~and
\NPB B331 541 1990 ;
V. Barger and T.G. Rizzo, \PRD D41 946 1990 ;
T.G. Rizzo, \IJMP A7 91 1992 .
\bibitem{rad2}
G.\ Altarelli, R.\ Barbieri, and S.\ Jadach, \NPB B369 3 1992 ;
G.\ Altarelli, R.\ Barbieri, and F.\ Caravaglios, CERN report
CERN-TH.6770/93 and G.\ Altarelli \etal , CERN report CERN-TH.6947/93.
\bibitem{tau}
T.G.\ Rizzo, \PRD D48 5286 1993 ;
X.-Y.\ Li and E.\ Ma, J.\ Phys. {\bf {G19}}, 1265, (1993).
\bibitem{old}
See, for example, J.L.\ Hewett and T.G.\ Rizzo, \PRD D45 161 1992 .
\bibitem{qw}
M.C.\ Noecker, B.P.\ Masterson, and C.E.\ Wieman, \PRL 61 310 1988  ;
S.A.\ Blundell, W.R.\ Johnson, and J.\ Sapirstein, \PRL 65 1411 1990 .
\bibitem{qwt}
W.J.\ Marciano and J.L.\ Rosner, \PRL 65 2963 1990 ~and
erratum \PRL 68 898 1992 ;
K.T.\ Mahanthappa and P.K.\ Mohapatra, \PRD D43 3093 1991 ~and erratum
\PRD D44 1616 1991 ;
T.G.\ Rizzo, University of Wisconsin report MAD/PH/604, 1990.


\end{thebibliography}
\end{document}